\documentclass[journal = aamick]{achemso}

\usepackage{amssymb}

\usepackage{graphicx}% Include figure files
\usepackage{dcolumn}% Align table columns on decimal point
\usepackage{bm}% bold math
\usepackage{upgreek}

\usepackage{epstopdf} 
\usepackage{titlecaps}
\Addlcwords{a an the of into onto on at in and or with but by as to}

\usepackage{etoolbox}
\makeatletter
\patchcmd{\acs@contact@details}{E}{Corresponding \,E}{}{}
\makeatother

\author{Wei-Hsiang Wang}
\affiliation{Department of Physics, National Taiwan Normal University, Taipei 116, Taiwan}
\alsoaffiliation{Department of Physics, National Sun Yat-sen University, Kaohsiung 804, Taiwan}
\author{Ching-Yang Pan}
\affiliation{Department of Physics, National Taiwan Normal University, Taipei 116, Taiwan}
\author{Chak-Ming Liu}
\affiliation{Department of Physics, National Taiwan Normal University, Taipei 116, Taiwan}
\author{Wen-Chin Lin}
\email{wclin@ntnu.edu.tw}
\affiliation{Department of Physics, National Taiwan Normal University, Taipei 116, Taiwan}
\author{Pei-hsun Jiang}
\email{pjiang@ntnu.edu.tw}
\affiliation{Department of Physics, National Taiwan Normal University, Taipei 116, Taiwan}

\title[]{Chirality-Induced Noncollinear Magnetization and Asymmetric Domain-Wall Propagation in Hydrogenated CoPd Thin Films}

\begin{document}

\begin{abstract}
Array-patterned CoPd-based heterostructures are created through e-beam lithography and plasma pretreatment that induces oxidation with depth gradient in the CoPd alloy films, breaking the central symmetry of the structure. Effects on the magnetic properties of the follow-up hydrogenation of the thin film are observed via magneto-optic Kerr effect microscopy. The system exhibits strong vertical and lateral antiferromagnetic coupling in the perpendicular component between the areas with and without plasma pretreatment, and asymmetric domain-wall propagation in the plasma-pretreated areas during magnetization reversal. These phenomenon exhibit evident magnetic chirality and can be interpreted with the Ruderman--Kittel--Kasuya--Yosida coupling and the Dzyaloshinskii--Moriya interaction (DMI). The sample processing demonstrated in this study allows easy incorporation of lithography techniques that can define areas with or without DMI to create intricate magnetic patterns on the sample, which provides an avenue towards more sophisticate control of canted spin textures in future spintronic devices.
\end{abstract}

\maketitle

\section{Introduction}

Synthetic magnets with assemblies of coupled magnetic elements that are microscopically tunable using external stimuli have attracted great attention for their potential to exhibit magnetic configurations with a well-defined hierarchy of energy levels of competing dipolar interactions among the elements \cite{Cowburn2000,Ross2002,Martin2003,Hrabec2020}. The synthetic magnets with controllable micron- to nanometer-sized chiral spin configurations are promising candidates for future information-processing devices with high energy efficiency \cite{Emori2013,Fert2013,Sampaio2013,Yu2012} and ultrahigh-density storage capability \cite{Fert2013,Rohart2013,Chatterjee2018}. To support dipolar interactions,  short-ranged quantum mechanical interactions provide suitable mechanisms to couple magnetic films in multilayers. Notable examples are the exchange bias between a ferromagnetic (FM) and an antiferromagnetic (AF) layer \cite{Nogues1999}, and the oscillatory Ruderman--Kittel--Kasuya--Yosida (RKKY) interaction between two magnetic layers separated by a normal metal \cite{Parkin1990}. Design of synthetic magnets for specific applications is thereby closely related to the ability to induce and tune the couplings between the magnetic elements.

Introduction of the Dzyaloshinskii--Moriya interaction (DMI) \cite{Dzyaloshinsky1958, Moriya1960} into the technology can provide an additional dimension for engineering the magnetic textures. DMI refers to the antisymmetric exchange at asymmetric interfaces owing to inversion symmetry breaking and spin--orbit coupling (SOC). This can occur in bulk materials with a composition gradient \cite{Robler2011,Kim2019}, as well as in multilayer magnetic films with naturally asymmetric interfaces, such as ferromagnets in contact with heavy metals \cite{Heide2008,Belabbes2016, Kikuchi2016,Jia2020}, graphene \cite{Belabbes2016,Yang2018}, or normal metals like Pd \cite{Pollard2020}, and ferrimagnetic garnets grown on nonmagnetic substrates \cite{Velez2019, Avci2019}. Among these experimental studies, the FM/graphene, FM/Pd, and ferrimagnet-based devices do not contain heavy metals to introduce strong SOC into the system. It is found that DMI along with the perpendicular magnetic anisotropy (PMA) induced in these systems is sufficient to stabilize Néel domain walls with a fixed chirality. These discoveries shed light on development of DMI spintronic devices with greater variability. 

On the other hand, controllable and reversible changes of magnetic properties of magnetic Pd alloy thin films induced by hydrogenation have recently been intensively studied for hydrogen sensing and magnetic engineering, with Pd serving as a catalyst to facilitate hydrogen absorption and desorption \cite{Remhof2008,Adams2011,Al-Mufachi2015,Silva2012}. 
In particular, the high hydrogen sorption capacity of CoPd alloys with a properly high concentration of Pd ($\geq$75\%) has been confirmed via measurements of the hydrogen solubility isotherms \cite{Zlotea2015}, the result of which can be further used for approximate calibration of the hydrogen concentration as a function of magnetic properties \cite{Chang2019}.
As hydrogenation is found to be able to manipulate noncollinear magnetic states by enhancing and tuning DMI \cite{Chen2021,Sandratskii2020,Yang2020,Hsu2018}, incorporation of the hydrogenation process should be considered for creating novel chiral magnets promoted by DMI.

In this study, we propose the possibility to develop CoPd-based heterostructures with DMI present in the system by employing oxidation with depth gradient and hydrogenation during the sample processing. Evident lateral AF coupling in the perpendicular component and asymmetric domain-wall propagation are observed from the processed CoPd alloy films via magneto-optic Kerr effect (MOKE) microscopy, demonstrating clear magnetic chirality in the system. The chiral behaviors are best interpreted with the existence of the RKKY coupling and the interlayer DMI in the oxidized-CoPd/Pd-rich/CoPd heterostructure.

\section{Material and Methods}

\begin{figure*}%[h]
\centering
\includegraphics[width=1\textwidth]{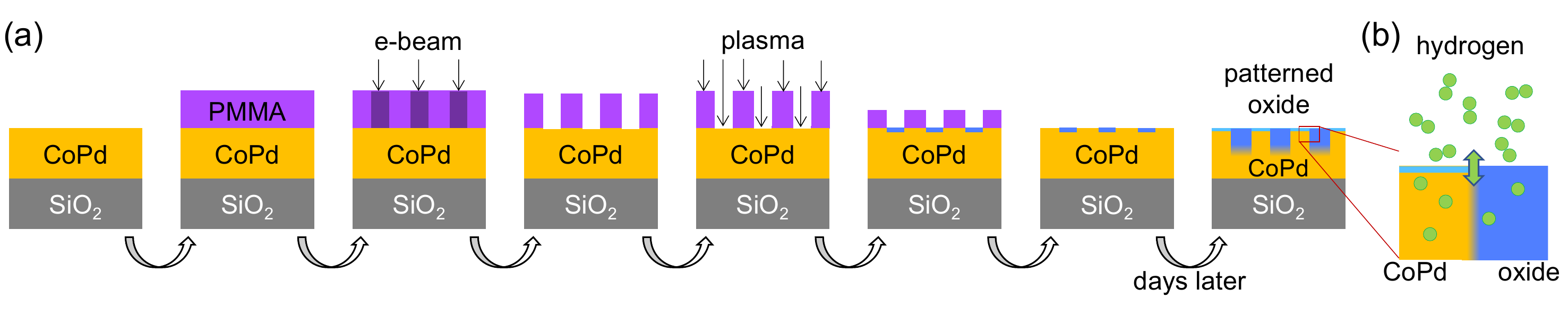}
\caption{(a) Patterning process that incorporates e-beam lithography and plasma treatment. Blue regions indicate partially oxidized regions. (b) Hydrogenation and dehydrogenation of the CoPd layer with patterned oxide.}
\label{fab}
\end{figure*}

The 8-nm-thick CoPd (25:75) alloy films were deposited on SiO$_2$/Si(001) substrates by e-beam heated co-evaporation of Co and Pd in an ultrahigh vacuum chamber with a base pressure of $3\times10^{-9}$ mbar. Two e-beam guns respectively for evaporation of Co and Pd were aligned at 45$^{\circ}$ to the normal of the substrate surface. This oblique deposition geometry allows uniaxial magnetic anisotropy (UMA) to be developed on the surface plane \cite{Chi2012}. The alloy composition and film thickness were controlled by the respective deposition rates of the elements, and were calibrated by X-ray photoelectron spectroscopy, atomic force microscopy, and transmission electron microscopy with energy dispersive X-ray spectroscopy \cite{Hsu2017}. 

Immediately following the film deposition was the surface processing using e-beam lithography and plasma treatment on the film, as illustrated in Fig.~\ref{fab}a. The CoPd film was spin-coated with a layer of polymethyl methacrylate (PMMA), which was later patterned through e-beam lithography to create a 7-by-7 array of $50 \times 40$ $\upmu$m$^2$ rectangular windows spaced 25 ${\upmu}$m apart along the longer side and 20 ${\upmu}$m apart along the shorter side of the rectangles, with the shorter side aligned with the magnetic easy axis of the as-deposited film (see Fig.~\ref{loops}a). The sample was then treated with 10.5-watt O$_2$ plasma (PDC-32G, Harrick Plasma) for 90 seconds with a fairly low flow rate of 0.1 cc/min of oxygen at a typical base pressure of 200 mTorr of the plasma chamber, followed by removal of PMMA. (According to the Langmuir probe measurement on a Harrick plasma cleaner of a similar model \cite{Nowak1990} and the plasma sheath model \cite{Panagopoulos1999}, the ion energy of the plasma is estimated to be $\sim$50 eV.) After that, the samples were stored in a vacuum desiccator (internal pressure $\sim0.15$ bar) to ensure mild and slow oxidation. The magnetic properties of the film in vacuum, under hydrogen exposure, and after evacuation of hydrogen, respectively, were inspected using MOKE microscopy 20 days after the plasma treatment. The hydrogenation and dehydrogenation processes are illustrated in Fig.~\ref{fab}b, with the hydrogen molecules dissociated into two unbound hydrogen atoms in CoPd because of the strong Pd--H interaction, thus weakening the H--H bond.

\section{Experimental Results and Discussion}
\subsection{Antiferromagnetic Coupling Observed in the Polar MOKE Hysteresis Loops}

\begin{figure*}%[h]
\centering
\includegraphics[width=0.95\textwidth]{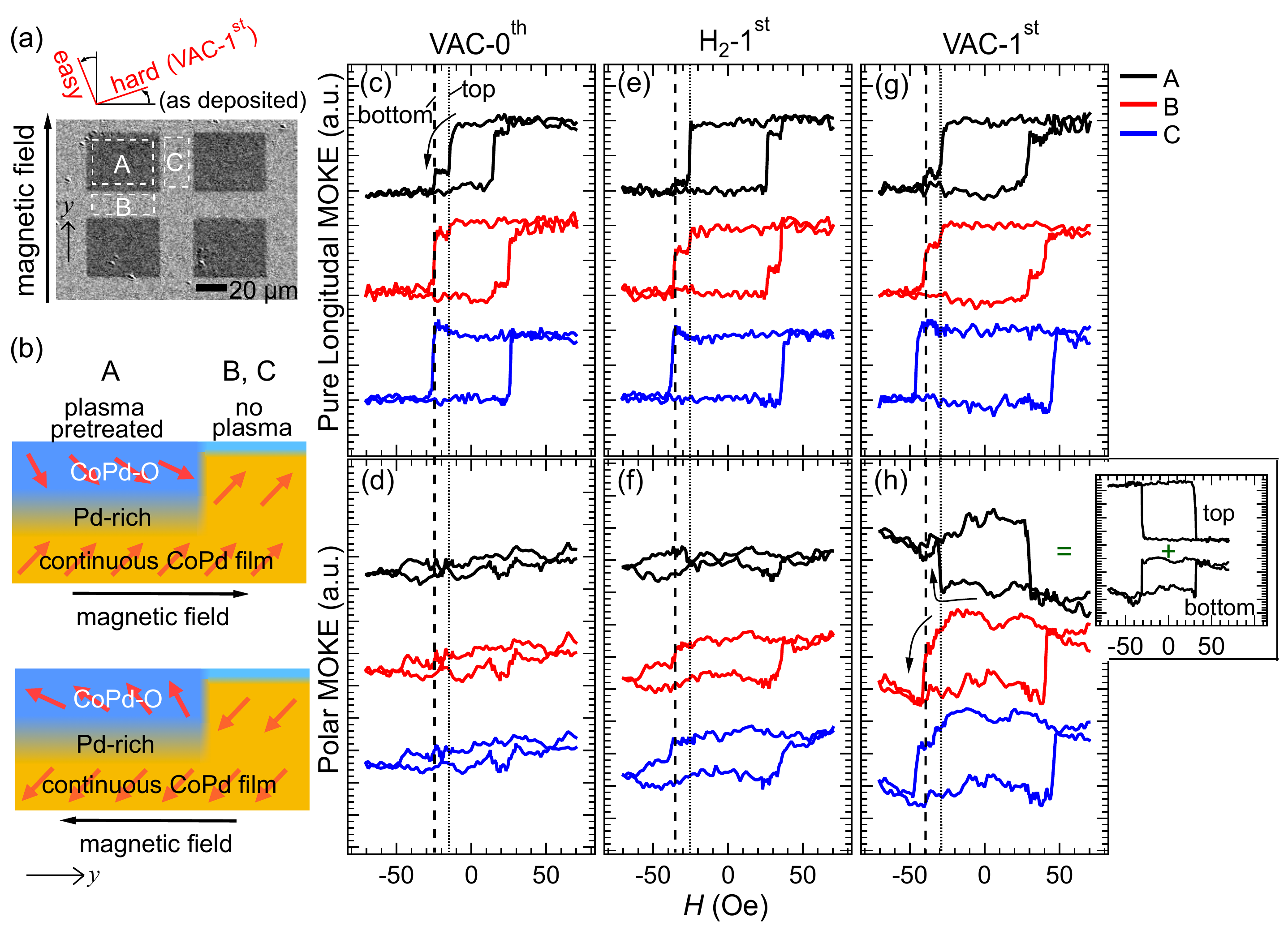}
\caption{(a) Polar MOKE image at the VAC-1\textsuperscript{st} stage at a magnetic field of 52 Oe along the easy axis of the as-deposited film. Area A is plasma-pretreated, whereas areas B and C are not. (b) Schematic diagram of the oxidized-CoPd/Pd-rich/CoPd multilayer in area A. Areas B and C have a much thinner layer of oxides on top. The red arrows indicate the magnetization directions of different areas at the VAC-1\textsuperscript{st} stage near saturation. (c)--(h) Pure longitudinal and polar MOKE loops at the VAC-0\textsuperscript{th} stage (c and d), H$_2$-1\textsuperscript{st} stage (e and f), and VAC-1\textsuperscript{st} stage (g and h), respectively. The VAC stages refer to the stages when the sample chamber is thoroughly evacuated, and the H$_2$ stage refers to the stage when the chamber is filled with 1-bar hydrogen. The black, red, and blue curves are measured from areas A, B, and C, respectively. The inset in (h) shows the signals from the top and the bottom layers that are decomposed from loop A in (h).}
\label{loops}
\end{figure*}

The magnetic field is applied along the easy axis of the as-deposited film in the experiment. Fig.~\ref{loops}a is a polar MOKE snapshot of the lithography- and plasma-pretreated CoPd film at a field of 52 Oe (retrieved from the measurement displayed in Fig.~\ref{loops}h) to exhibit high contrast of magnetic domains for illustrating the respective areas from which the MOKE signals are measured. Area A was exposed to plasma during the plasma pretreatment, whereas areas B and C were protected from plasma by the PMMA layer. The sample then undergoes three different ambient conditions in which the MOKE measurements are conducted, respectively. VAC-0\textsuperscript{th} (labeled on top for Figs.~\ref{loops}c and \ref{loops}d) refers to a stage when the MOKE microscope chamber hosting the sample is pumped to a vacuum of $5\times10^{-5}$ mbar for the first time, H$_2$-1\textsuperscript{st} (Figs.~\ref{loops}e and \ref{loops}f) refers to the stage when the chamber is filled with 1-bar hydrogen after VAC-0\textsuperscript{th}, and VAC-1\textsuperscript{st} (Figs.~\ref{loops}g and \ref{loops}h) corresponds to a stage when the chamber is pumped to $5\times10^{-5}$ mbar again after H$_2$-1\textsuperscript{st}. By addition and subtraction of the MOKE images from the opposite directions of oblique incidence, a pure polar image and a pure longitudinal image can be obtained, respectively \cite{Soldatov2017}. The hysteresis loops in the top figures (Figs.~\ref{loops}c, \ref{loops}e, and \ref{loops}g) are acquired using pure longitudinal MOKE (LMOKE), whereas the bottom ones (Figs.~\ref{loops}d, \ref{loops}f, and \ref{loops}h) are obtained using pure polar MOKE (PMOKE). The black curves are measured from area A indicated in Fig.~\ref{loops}a, and the red and blue curves are from areas B and C, respectively. PMOKE measurements detect magnetization vectors that are perpendicular to the sample surface. As PMOKE signals are inherently optically stronger than the LMOKE signals by an order of magnitude \cite{Kuch2015}, the top LMOKE plots are presented in a vertical scale different from that of the bottom PMOKE plots for visual clarity.

In Fig.~\ref{loops}c, loop A exhibits magnetization reversal with two steps observed on both the positive and the negative field sides, with the steps marked by the dotted line and dashed line, respectively. This indicates the existence of two different FM sources in area A. (Measurement of a minor hysteresis loop at the first step exhibits no exchange bias (see the Supporting Information), indicating that there is no observable coupling between the two FM sources \cite{Grolier1993, Liu2003}.) This two-step reversal is absent from loop C, which implies modification of the film properties induced by plasma pretreatment on area A. It is known from our previous studies that plasma pretreatment triggers faster surface oxidation of the CoPd film, and oxidation leads to reduction in the coercive field $H_\mathrm{c}$ \cite{Wang2021}. Therefore, the first and the second steps of the reversal in area A  are attributed  to the top partially-oxidized layer and the bottom CoPd layer, respectively, as illustrated in Fig.~\ref{loops}b. As plasma treatment and the follow-up oxidation can induce segregation and migration of some Co atoms in the film to form Co oxides on the surface, a thin layer of Pd-rich phase is established above the bottom CoPd layer underneath the oxides \cite{Aguilera-Granja2006, Hsu2017, Wang2020}. (The X-ray photoelectron spectroscopy depth profile of the film is provided in the Supporting Information.) Area C is much less oxidized than area A in the sense that the former has a much lower Co-oxide concentration and a much smaller oxidation depth into the film, exhibiting only one dominating reversal process ascribed to the bottom CoPd layer. Area B is as minutely oxidized as area C, yet it shows a small step corresponding to the magnetization reversal of the top layer. This indicates the significance of the magnetostatic interaction along the easy axis as area B is sandwiched by two plasma-pretreated rectangles along the easy axis, which makes area B more likely to be influenced by area A than is area C. The PMOKE measurements in Fig.~\ref{loops}d, on the other hand, exhibit very small signals.

The sample is then immersed in hydrogen (i.e., the H$_2$-1\textsuperscript{st} stage), and the MOKE measurements are conducted after the magnetic properties of the sample are stabilized. In hydrogen, $H_\mathrm{c}$ from the pure LMOKE measurements increases for all the areas as shown in Fig.~\ref{loops}e, with loop C showing a slightly larger  $H_\mathrm{c}$ than the other two. The increase in $H_\mathrm{c}$ has also been observed in our previous studies of hydrogenated CoPd thin films \cite{Lin2015,Lin2016}, which may be attributed to the increase in the magnetocrystalline anisotropy caused by hydrogen-induced modulation of the electronic structures. It is also noticed from Fig.~\ref{loops}f that the PMOKE signals begin to develop upon hydrogen exposure when the external field is applied solely in plane. Afterward, the sample chamber is pumped down again (i.e., VAC-1\textsuperscript{st}) to perform dehydrogenation, and the results are shown in Figs.~\ref{loops}g and \ref{loops}h. (The magnetic axes are slightly rotated upon hydrogen exposure, as indicated with the red axes in Fig.~\ref{loops}a. This phenomenon will be furthered discussed in Sec.~\ref{asymwall}.) It is found that all $H_\mathrm{c}$'s become even larger, with $H_\mathrm{c}$ of loop C increases the most. The initial pure LMOKE loops seen at VAC-0\textsuperscript{th} (Fig.~\ref{loops}c) can not be recovered by hydrogen evacuation. The most intriguing phenomena observed at this stage are the strong PMOKE signals shown in Fig.~\ref{loops}h. Loop A, in particular, exhibits a reversed polarity with respect to the other two and with respect to all the PMOKE loops at H$_2$-1\textsuperscript{st} (Fig.~\ref{loops}f). The reversed loop may imply the presence of AF coupling between the top partially-oxidized layer of area A and its adjacent regions \cite{Roy2005}, as the perpendicular magnetization of area A is opposite to all the other areas most of the time, except during magnetization reversal when area A switches magnetization at a lower in-plane field (owing to a lower $H_\mathrm{c}$) than the other areas, acquiring the same perpendicular magnetization direction as those of areas B and C before the latter two switch magnetization. It is therefore speculated that the top partially-oxidized layer of area A is AF coupled with its bottom CoPd layer, as opposed to the top portion of areas B and C remaining FM coupled with the continuous bottom layer. The magnetization at a high positive in-plane field (e.g., 52 Oe), is illustrated with red arrows in the top panel of Fig.~\ref{loops}b, whereas the bottom panel describes the magnetization at a high negative in-plane field. The orientation of local magnetization in the oxidized CoPd layer bears some helicity from side to side because of DMI \cite{Vedmedenko2019}, which will be further discussed in Sec.~\ref{asymwall}.

Similar trends of small oscillations are observed in all three loops in Fig.~\ref{loops}h, and are attributed to the continuous bottom layer that is present in all three areas. Loops B and C are contributed mostly by the CoPd layer, because the top oxidized crust is too thin. By comparing the small oscillations among the loops, loop A can thereby be decomposed into the signal from the top partially-oxidized layer and that from the bottom continuous CoPd, as illustrated in the inset. The bottom component of loop A thus resembles loops B and C, whereas the top component has a magnitude much larger than the PMOKE signals from areas B and C. This demonstrates that, along with the reversed polarity, plasma pretreatment that leads to significant hydrogen trapping by the resulting alloy vacancies also gives rise to a considerable enhancement of the magnitude of the perpendicular magnetization in the top oxidized CoPd layer.

The key features of the hysteresis loops observed at H$_2$-1\textsuperscript{st} and VAC-1\textsuperscript{st} can be reproduced for several cycles of refilling the sample chamber with hydrogen and evacuating it, including the larger $H_\mathrm{c}$ at each subsequent VAC stage than at its previous H$_2$ stage, the stronger PMOKE signals at the VAC stages than at the H$_2$ stages, and the disappearance of AF coupling at the H$_2$ stages and its reappearance (demonstrated by the reversed loop A) at the VAC stages observed in PMOKE. The two-step reversal in LMOKE at all stages also persist for several cycles. 

\subsection{Bordering of the Plasma-Pretreated Areas by Edge Domains Observed in the Polar MOKE Images}

\begin{figure*}%[h]
\centering
\includegraphics[width=1\textwidth]{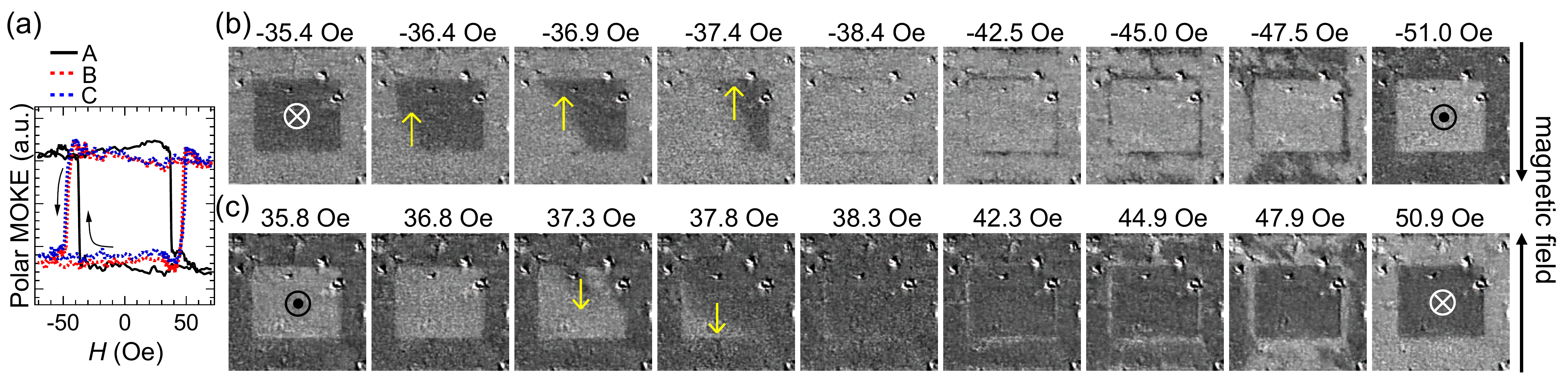}
\caption{(a) PMOKE hysteresis loops of areas A, B, and C, respectively, at another subsequent VAC stage. (b) PMOKE images demonstrating the evolution of the magnetic domains inside and outside a rectangle upon magnetization reversal as the magnetic filed is varied from positive to negative fields. (c) PMOKE images of the same rectangle upon magnetization reversal as the magnetic filed is varied from negative to positive fields. The white and black notations at the center of the rectangle indicate the directions of the perpendicular magnetization inside the rectangle. The yellow arrows on a domain wall indicate the directions of the wall propagation.}
\label{asym}
\end{figure*}

Strong PMOKE signals shown in Fig.~\ref{loops}h indicate prominent perpendicular magnetization after hydrogen exposure and evacuation. It has been found in our previous study that the magnetic anisotropy is dominated by PMA for CoPd films with thicknesses of $\lesssim$10 nm \cite{Chang2017}, which can be attributed to the tetragonal $L$1$_0$ ordering in the CoPd alloy \cite{Sedrpooshan2018,Komninou2021,Vivas2016} even on the short-range scale \cite{Kim1997,Vivas2016}, and to the interfacial anisotropy or strain \cite{Okabayashi2021,Chappert1995,Lin2000,Peng2015,BeikMohammadi2019,Liu2019}. Oxidation diminishes PMA \cite{Hsu2017}, but hydrogenation afterward can enhance PMA with proper hydrogen concentrations \cite{Lin2013a, Lin2015a, Lin2016,Lin2013,Lin2016a,Klyukin2020}. The perpendicular magnetization under in-plane fields after hydrogenation can be attributed to the PMA and the long-range coupling \cite{Lin2015} induced by hydrogenation that modulates the electronic structure and orbital moments of the alloy atoms. Indirect exchange coupling of  Co magnetic moments is established via the conduction electrons of Pd between Co atoms, and the uptake of hydrogen changes the electronic structure of Pd and accordingly the interaction. Because PMOKE loops reveal several unusual behaviors after hydrogen evacuation, Fig.~\ref{asym} is arranged to focus on the evolution of the PMOKE images at a subsequent VAC stage for further investigation. The resolution of the MOKE images at the magnification used (200$\times$) is 1.29 ${\upmu}$m with our high-resolution CCD sensor for detection of the LED light reflected from the sample. Fig.~\ref{asym}a is from a  VAC stage different from Fig.~\ref{loops}h, but replicates the key features including the reversing of loop A, demonstrating the reproducibility of these measurement results. As the field is swept from positive (up) to negative (down), loop A stays at the lower bound, and loops B and C stay at the upper bounds. Then, magnetization reversal takes place, with loop A rising up and loops B and C dropping down. The detailed magnetic domain dynamics of the reversal procedure is displayed in the series of MOKE images shown in Fig.~\ref{asym}b zoomed in at a single rectangle for a close-up inspection. As the field becomes more negative from $-35.4$ to $-51.0$ Oe, the rectangle turns from dark to light, and the outside areas turn from light to dark. Magnetization reversal also occurs when the field is swept from negative to positive, starting around 35.8 Oe, with loop A dropping down and loops B and C rising up. This corresponds to the image series  shown in Fig.~\ref{asym}c,  as the rectangle turns from light to dark, and the outside areas turn from dark to light. AF coupling is thereby unambiguously observed in both field sweep directions through the opposite magnetization between inside and outside the rectangle. At fields where areas A, B, and C exhibit the same gray shades, that is, from $-38.4$ to $-42.5$ Oe and from 38.3 to 42.3 Oe when area A just finishes magnetization reversal and before areas B and C start to reverse, a clear trace of AF coupling can still be detected at the edges of the rectangle. The light rectangle in Fig.~\ref{asym}b is bordered with a dark perimeter,  and the dark rectangle in Fig.~\ref{asym}c with a light perimeter, demonstrating the strong inclination for the outside areas to display a magnetization opposite to the inside area as they confront the newly reversed rectangle at the boundaries. 

The bordering of the rectangles by the edge domains with opposite perpendicular magnetization points to a strong \textit{lateral} AF coupling between the areas inside and outside the rectangles, in addition to the vertical AF coupling between the top and bottom layers that also exists in the system. Strong coupling of laterally adjacent magnetic domains has recently drawn great attention for its potential to advance realization and engineering of two-dimensional networks of magnetic elements for constructing planar logic gates and memory devices. Luo \textit{et al.}~recently discovered strong chiral lateral coupling between adjacent in-plane and out-of-plane magnetic structures in array-patterned Pt/Co/AlO$_x$ with strong DMI present in the system \cite{Luo2019}. When considering the total magnetization, the coupling in our system also satisfies some chirality (see Fig.~\ref{loops}b)---magnetization pointing down right ($\searrow$) inside the rectangle at a high positive in-plane field induces an outside rim of magnetization pointing up right ($\nearrow$) , whereas magnetization pointing up left ($\nwarrow$) inside the rectangle at a high negative in-plane field induces an outside magnetization pointing down left ($\swarrow$) (``right/left'' means the $+y$/$-y$ direction). DMI may also be present in our system to support the magnetic interaction and spin canting with helicity \cite{Vedmedenko2019}, as will be further discussed in Sec.~\ref{asymwall}.

It has recently been known that hydrogenation can induce and tailor AF coupling in several materials \cite{Hjorvarsson1997,SanchezMarcos2007,Shashikala2009,Jani2021}, but hydrogenation-induced AF coupling in CoPd is reported for the first time. Exchange coupling between two FM systems may be manipulated by hydrogenation via the change of the electronic structure and the spin polarization of the subnanostructured Pd present between the two FMs, whether these substances are constructed in a multilayer structure or in an alloy, as predicted by the RKKY interaction \cite{Baltensperger1990, Klose1997, Leiner2003, Lin2015, Gobler2019,Gobler2021}. The Pd-rich layer between the two FM layers in our system is still a CoPd alloy but with a higher concentration of Pd. While Co subnanoclusters are coupled via RKKY exchange interaction mediated by hydrogen-modulated conduction electrons of the Pd matrix in the alloy, the transition between the bottom CoPd and the Pd-rich layer also bears a certain resemblance to the FM/Pd interface \cite{Gobler2021}. Therefore, the well-established RKKY interaction in the Pd-rich layer further couples the bottom CoPd layer and the top partially-oxidized CoPd layer in our system. With RKKY bilinear interaction, the two AF-coupled layers should have favored antiparallel orientation in their magnetizations. Instead, canted coupling is present in our system, which may be ascribed to an interplay between RKKY and PMA, or to the DMI between the two FM layers.

To observe the desired AF coupling effects, the concentration of hydrogen in the film is critical. Evacuation of hydrogen gas is always needed after hydrogen exposure to achieve a low enough concentration of hydrogen in the film for it to exhibit the AF coupling effects discovered in the PMOKE measurements. The persistence of the remnant hydrogen inside the film even after thorough evacuation can be accounted for by the trapping effect of alloy vacancies on hydrogen. Since oxidation creates defects, it is conjectured that a small amount of hydrogen is trapped in the defects or vacancies of the alloy film, not being able to be removed upon evacuation of the vacuum chamber unless assisted with thermodesorption via sample heating \cite{You2013,Kammenzind2018}. The persistence of the remnant hydrogen is also dependent on the ratio of Pd to Co in the alloy film, according to our previous study \cite{Lin2016}. As Pd acts as the catalyst to facilitate hydrogen absorption in metal-hydrides, films with higher percentages of Pd are more likely to lose the ability to regain the original hysteresis loop in initial vacuum though hydrogen evacuation. In short, a proper amount of oxidation-induced defects and a higher ratio of Pd are critical to observe the remarkable phenomenon of AF coupling in the CoPd film.

\subsection{Asymmetric Domain-Wall Propagation Observed in the Polar MOKE Images}\label{asymwall} 

Another important signature observed in Figs.~\ref{asym}b and \ref{asym}c is the asymmetric domain-wall propagation 
inside the rectangle. This asymmetric reversal behavior, like the AF coupling effect, can only be observed in PMOKE at the subsequent VAC stages. 
Domain walls observed in the MOKE images are found to be misaligned with the easy axis of the as-deposited film, which can be attributed to the rotation of the easy axis induced by hydrogenation \cite{Wang2021a} (see the Supporting Information for the angular dependence of the magnetic remanence of the as-deposited film and of the film at the subsequent VAC stage). In Fig.~\ref{asym}b as the field increases in the negative direction, the tilted domain wall inside the rectangle propagates from bottom to top (indicated by the yellow arrows). However, in Fig.~\ref{asym}c as the field increases in the positive direction, the wall propagates in the opposite direction from top to bottom. The direction switching of the wall propagation clearly reveals symmetry breaking and chiral interaction in our system. As illustrated in Fig.~\ref{loops}b, local magnetization in the top partially-oxidized CoPd layer in area A most likely exhibits a varying tilt angle from side to side  that is determined by the boundary condition of a microstructured magnetic thin film with DMI \cite{Rohart2013,Pizzini2014,Garcia-Sanchez2014}. Application of the in-plane magnetic field plays a significant role to decide the dominating site where the magnetization reversal starts---the edge magnetization with a polarization closer to that of the in-plane field should be affected by the field more efficiently and thereby leads the reversal \cite{Han2016}.  If the field is applied in the $+y$ ($-y$) direction, the magnetization configuration changes from the bottom (top) panel in Fig.~\ref{loops}b to the top (bottom) panel during the reversal. As the magnetization orientation (indicated by the red arrows) on the $+y$ ($-y$) side in the bottom panel is \textit{less antiparallel} to the applied $+y$ ($-y$) field, the domain-wall motion starts from there. This concludes a direction switching of the wall propagation under an opposite in-plane field, 
from an upward prorogation under a downward in-plane field as shown in Fig.~\ref{asym}b to a downward prorogation under an upward in-plane field in Fig.~\ref{asym}c. 

Asymmetric wall propagation has also been observed in Pt/Co/AlO$_x$ and Pt/Co/Ir thin films with \textit{intralayer} DMI present in the systems \cite{Han2016, Pizzini2014}. Monte Carlo simulations of a Pt/CoFeB/Pt/Ru/Pt/Co/Pt thin film, on the other hand, evidence an emergence of the asymmetric magnetization reversal as a result of the competition between the \textit{interlayer} DMI and the RKKY coupling \cite{Fernandez-Pacheco2019}. The interlayer DMI refers to the antisymmetric coupling between two different FM layers mediated by a thin spacer of a normal paramagnetic metal with relativistic spin--orbit interaction \cite{VanWaeyenberge2019,Fernandez-Pacheco2019,Han2019,Vedmedenko2019}. The asymmetric wall propagation in our system may be interpreted with the interlayer DMI, as the Pd-rich layer illustrated in Fig.~\ref{loops}b can serve as the mediating spacer between the top and the bottom FM layers in our system. While intralayer DMI may also exist in our system, it may not suffice to explain the asymmetric wall propagation that takes place in the absence of a perpendicular magnetic field, because observations of the effects of the intralayer DMI require the simultaneous application of perpendicular magnetic fields for nucleation \cite{Han2016, Pizzini2014, Jue2016,Vanatka2015,Lau2016}, in contrast to our experiments. For interlayer exchange coupling between a soft layer (the partially-oxidized CoPd layer in our case) and a harder PMA layer (the bottom CoPd layer in our case), the harder layer serves to assist nucleation even without an external perpendicular magnetic field \cite{Fernandez-Pacheco2019}. (Additional information of the asymmetric wall propagation in different rectangles in the array is provided in the Supporting Information.)

The contribution of the oxidized-CoPd/Pd-rich/CoPd interfaces to the effective DMI is a subject of interest, as the absence of strong SOC in this system calls for a new microscopic mechanism. Strong DMI contributions have been predicted for Co/O interfaces \cite{Freimuth2014,Yang2018,Boulle2016,Chaves2019}. Moreover, asymmetry in magnetic domains was recently observed in Co/Pd multilayers, and was explained with numerical models of an interlayer DMI \cite{Pollard2020}. SOC is allowed for electrons near the Co/Pd interface, and the RKKY interaction mediated by the spin--orbit coupled electrons leads to the perpendicular component of the interlayer DMI. Presence of DMI is thereby also expected in our oxidized-CoPd/Pd-rich/CoPd system even without the presence of heavy metals to harbor intrinsic strong SOC. 

\section{Conclusions}

In summary,  we report on observation of chiral magnetic properties of dilutely hydrogenated CoPd thin films that are array-patterned via e-beam lithography and oxygen plasma. The two steps detected in the magnetization reversal in the pure LMOKE hysteresis loops are attributed to the two FM layers of the oxidized-CoPd/Pd-rich/CoPd heterostructure created in the plasma-pretreated areas. After hydrogenation and dehydrogenation of the the sample with remnant hydrogen being persistently trapped in the alloy vacancies of the film, strong vertical and lateral AF coupling in the perpendicular component are observed between the plasma-pretreated areas and the untreated ones through the PMOKE loops and the domain evolution shown in the PMOKE images. The PMOKE images also demonstrate an intriguing asymmetric domain-wall propagation in the plasma-pretreated area, pointing to a possible competition between the interlayer DMI and the RKKY coupling present in our system. Plasma-induced oxidation with depth gradient followed by proper hydrogenation is therefore considered a practicable tool to develop a heterostructure in a magnetic film with broken central symmetry that allows DMI to exist in the system. Moreover, our sample processing allows easy incorporation of lithography techniques that can define areas with or without DMI to create intricate magnetic patterns on the sample. These techniques can be applied in fabrication of future spintronic devices with sophisticate control of canted spin textures.

\section*{Supporting Information}

Measurement of the minor hysteresis loop in a plasma-pretreated rectangle in the VAC-0\textsuperscript{th} stage, X-ray photoelectron spectroscopy depth profiling, angular dependence measurements, and additional information of the asymmetric wall propagation of the rectangles in the 7-by-7 array.

\section*{Acknowledgments}

We acknowledge the group of Prof.~Hsiang-Chih Chiu for assisting us with the plasma source. This study is sponsored by the Ministry of Science and Technology of Taiwan under Grants Nos.~MOST 109-2112-M-003-009, MOST 110-2112-M-003-019, and MOST 110-2112-M-003-014.

\bibliography{H2CoPd}
\newpage

\end{document}